\documentclass[useAMS,usenatbib]{mn2e}
\usepackage{psfig}

\usepackage{times}

\def\lesssim{{_ <\atop{^\sim}}}

\def\fb{\mbox{$f_{\rm gal}$}}

\def\rd{\mbox{h$_d$}}

\def\hB{\mbox{$R_{B}$}}
\def\hV{\mbox{$R_{V}$}}

\def\hs{\mbox{$R_{*}$}}

\def\rh{\mbox{$R_{\rm h}$}}
\def\md{\mbox{$M_{\rm bar}$}}

\def\ms{\mbox{$M_{\rm *}$}}
\def\mh{\mbox{$M_{\rm h}$}}
\def\msun{\mbox{M$_\odot$}}

\def\LCDM{\mbox{$\Lambda$CDM}}

\def\apj{\mbox{ApJ}}
\def\mnras{\mbox{MNRAS}}
\def\apjl{\mbox{ApJL}}

\def\aj{\mbox{AJ}}

\def\spose#1{\hbox to 0pt{#1\hss}}
\newcommand\lsim{\mathrel{\spose{\lower 3pt\hbox{$\mathchar"218$}}
     \raise 2.0pt\hbox{$\mathchar"13C$}}}
\newcommand\gsim{\mathrel{\spose{\lower 3pt\hbox{$\mathchar"218$}}
     \raise 2.0pt\hbox{$\mathchar"13E$}}}

\title[Size evolution of galaxy discs ]
{The size evolution of galaxy discs formed within $\Lambda$ Cold Dark Matter haloes}

\author[Firmani \& Avila-Reese]
{C. Firmani$^{1,2}$\thanks{E--mail: firmani@merate.mi.astro.it} 
and V. Avila--Reese$^{2}$\\
$^{1}$Osservatorio Astronomico di Brera, via E.Bianchi 46, I-23807
Merate, Italy\\
$^{2}$Instituto de Astronom\'{\i}a, Universidad Nacional Aut\'onoma de M\'exico,
A.P. 70-264, 04510, M\'exico, D.F.}

\begin{document}


\pagerange{\pageref{firstpage}--\pageref{lastpage}} \pubyear{2002}

\maketitle

\label{firstpage}

\begin{abstract}
By means of galaxy evolutionary models, we explore the direct consequences 
of the $\Lambda$ Cold Dark Matter (\LCDM) cosmogony on the size evolution of galactic discs,
avoiding intentionally the introduction of intermediate (uncertain) astrophysical 
processes. Based on the shape of the rotation curves and guided by a simplicity criterion, 
we adopt an average galaxy mass baryon fraction $\fb=0.03$. In order to study general 
behaviors, only models with the average initial conditions are analyzed. 
The stellar and $B$-band effective radii, \hs\ and \hB, of individual galaxies grow 
significantly with time (inside-out disc formation) with laws that are weakly dependent on 
stellar mass, \ms, or luminosity, $L_B$. However, the change of \hs\ with $z$ at a 
{\it fixed} \ms\ is slow; for $z\leq 2.5$, \hs(\ms=constant)$\propto (1+z)^{-0.4}$ for a large range 
of masses. On the other hand, the change of \hB\ with $z$ at a {\it fixed} $L_B$ is strong and it
resembles the \hB\ decreasing law of the individual models; roughly 
\hB($L_B$=constant)$\propto (1+z)^{-0.85}$ for $z\lesssim 0.75$, and $\propto (1+z)^{-1.1}$ 
for higher $z'$s. 
We find also that at $z=0$, $\hs\propto \ms^{0.38}$ and $\hB\propto L_B^{0.40}$, remaining
the slopes of these relations practically the same up to $z\approx 3$. Our model predictions
are in reasonable agreement with observational inferences on the typical radius change with $z$ 
of late-type galaxies more luminous (massive) than high values imposed by the selection 
effects. The  models seem also to be consistent, within the large scatter, with the 
\hB\ and $L_B$ values obtained from small (non complete) samples of sub-$L_*$ late-type 
galaxies with available rest-frame photometric information at different $z'$s.
The properties and evolution of the \LCDM\ haloes seem to be the main drivers of 
galaxy disk size evolution.
Nevertheless, the models reveal a potential difficulty in explaining the observed steepening
of the \hB-$L_B$ relation with respect to the \hs-\ms\ one, an effect related to the 
well established color-magnitude relation.

\end{abstract}


\section{Introduction}

The completion of multi-wavelength deep field surveys during the last years has enormously 
benefited the study of galaxy evolution. The analysis of galaxies with redshift known from 
these surveys provides valuable information about the changes with cosmic time of galaxy 
population properties. However, since the evolution of individual galaxies is not directly 
observable and because of the strong selection effects at higher redshifts, a direct 
interpretation of the observations is not an easy task.  An adequate comparison of 
theoretical models with observations helps largely in 
this undertaking. Here we will present evolutionary models of disc galaxy formation and 
evolution inside growing $\Lambda$ Cold Dark Matter (\LCDM) haloes with the aim to show 
predictions (i) on size evolution of {\it individual} galaxies, and (ii) on the typical size change 
with redshift $z$ of galaxies of a fixed stellar mass, \ms, or luminosity, $L$.  The latter 
results are those that can be compared with the information provided by observations.  

The qualitative description of disc galaxy formation in \LCDM\ haloes is based on the idea that 
both dark and baryonic matter acquire angular momentum by tidal torques while the perturbation 
is in its linear regime. The gas within the virialized halo cools and falls to its center until 
it attains centrifugal equilibrium forming a disc (White \& Rees 1978). According 
to the spherical collapse model, an overdense region 
of mass \mh\ virializes with a radius \rh\ at the epoch $z$ when its cumulative density overcomes 
$\rho_{\rm bg}(z)\times \Delta_c(z)$, where $\rho_{\rm bg}(z)$ is the background density at $z$
and $\Delta_c$ is a critical value at each epoch $z$ that depends on the cosmological model 
(Bryan \& Norman 1998). It follows then that 
\begin{equation}
\rh = 1.63 \left(\frac{\mh}{\msun}\right)^{1/3} \left(\frac{1}{(\Delta_c(z)/2)\ H(z)^2}\right)^{1/3}  \rm kpc, 
\label{radvir}
\end{equation}
where $H(z)$ is the Hubble parameter. If the disc galaxy radius is proportional to \rh, then 
a significant decrease of sizes with $z$ at a fixed halo mass is expected (e.g., Mo, Mao, \& 
White 1998; Bouwens \& Silk 2002). For haloes treated as truncated isothermal spheres and 
assuming that an exponential disc is formed under detailed angular momentum conservation, 
it indeed follows that $\rd\propto \lambda\rh$ (e.g., Fall \& Efstathiou 1980; Mo et al.  
1998), where \rd\ is the disc scale length and $\lambda$ is the halo spin parameter 
(Peebles 1969). Taking into account a Navarro-Frenk-White (NFW) density profile and 
including the adiabatic halo contraction due to disc formation, the latter dependence 
becomes $\rd\propto \lambda\rh \ f_c^{-1/2}\ f_R(\lambda,c,\fb),$ where $c$ is the 
concentration parameter of the NFW profile, $f_c$ is a shape-function that increases with 
$c$, $\fb\equiv \md/\mh$ is the mass galaxy baryon fraction, and $f_R$ is a function 
that takes into account the NFW halo adiabatic contraction for an exponential baryonic disc 
(Mo et al. 1998). Somerville et al. (2008) have shown that a further reduction of the 
change of galaxy radius with $z$ at a fixed mass is obtained considering the fact that 
$c$ decreases roughly as $(1+z)^{-1}$. These authors concluded that the observed weak 
change with $z$ of typical stellar radii of disc galaxies of similar stellar masses at 
all redshifts agrees with the model predictions under the key assumption that $\lambda$ 
and \fb\ do not change with $z$.

The models based on the Mo et al. (1998) approach (e.g., those of Somerville et al. 2008)
actually refer to a 'static' (instantaneous) population of disc galaxies at a given $z$; 
{\it the models do not follow actually the individual and local evolution of the halo-galaxy system.} 
Besides, such models consider only baryonic discs and assume they have an exponential 
mass surface density distribution. 
The complex evolutionary processes of gas infall and local gas transformation into stars, with the 
consequent prediction of gas, stellar, and luminous disc properties, are beyond the scope of the 
mentioned models.  The cosmological N-body/hydrodynamical simulations are on the other extreme 
of complexity. These simulations use to report the so-called 'angular momentum 
catastrophe' with the consequent formation of too small and compact discs (Navarro \& White 
1994). Such a problem seems to have been largely ameliorated in recent simulations of small 
volumes with a very high resolution and an appropriate tuning of sub-grid physics 
(e.g., Abadi et al. 2003; Governato et al. 2004,2007; Robertson et al. 2004;
 Zavala, Okamoto \& Frenk 2008; Scannapieco et al. 2008).  
Even so, the full numerical simulations do not yet allow to predict in detail the evolution 
of a whole population of well resolved galaxies. 

From the observational side, the study of galaxy size evolution is hampered by the 
flux and surface brightness (SB) limits, which introduce incompleteness and 
selection effects in the photometric samples, particularly for those at 
higher redshifts. Thus, the observational inferences are yet controversial 
(see for recent reviews Cameron \& Driver 2007; Elmegreen et al. 2007).  
Because magnitude and size correlate strongly, the determination
of the average rest-frame surface brightnesses (SB) as a function of $z$ has been used 
commonly to parametrize the evolution of galaxy sizes. In this analysis is crucial
the choice of the selection and completeness functions to be applied to each $z$ bin. 
In early works, an increase of the mean rest-frame $B-$band SB 
of $\sim 1$ mag from $z\approx 0$ to $z\sim 1$ was reported for late-type galaxies (Schade et al.
1995; Roche et al. 1998). Later on, the use of stricter completeness functions, defined for 
the highest $z$ bins and imposed to the lower-$z$ bins, showed no detectable mean SB evolution 
(Simard et al. 1999; Ravindranath et al. 2004). More recently, larger galaxy samples were 
used, and the completeness functions were estimated independently for each $z$ bin by means 
of artificial galaxy simulations. As a result, a non-negligible SB or radius (at a given 
luminosity) evolution for luminous late-type galaxies was inferred (e.g., Bouwens \& Silk 2002;
Bouwens et al. 2004; Barden et al. 2005; Trujillo et al. 2006; Cameron \& Driver 2007; 
Franx et al. 2008). 
Some of these samples extend up to $z\sim 3$ or even $z\sim 6$. However, as higher is $z$,  
(i) only the most luminous and highest SB galaxies of the epoch are observed,  and (ii) 
the determination of the galaxy type becomes very uncertain.

It is important to have in mind that, likely, both radius and luminosity evolve. 
Therefore, is not straightforward to decipher from the sample mean SB change with
$z$ how is the size evolution of {\it individual} discs.  Besides, the SB 
evolution inferences happen to be different in different rest-frame bands. Additional 
kinematic information (rotation velocities) certainly may help on this undertaken 
(see e.g., Mao, Mo \& White 1998).

Given the current status of the theory and observations mentioned above,
the following natural questions arise. Do the direct implications of the \LCDM\ 
cosmogony on galactic disc evolution agree with the observational inferences? 
Could galaxy evolutionary models shed light on the connection between the individual 
evolution of discs and the sample properties inferred from photometric observations?

With the aim to answer these questions, we will use here the {\it semi-numerical} evolutionary 
approach: the assembling of haloes and the discs within them is (i) followed 
self-consistently (for the discs we use hydrodynamic techniques and take 
into account realistic descriptions for the relevant astrophysical 
processes, e.g., star formation, turbulence and its dissipation, 
stellar dynamics, gravitational instability, etc.), and (ii) for a large range of 
initial conditions (Avila-Reese et al. 1998; Firmani \& Avila-Reese 2000 
--hereafter FA2000; Avila-Reese \& Firmani 2000,2001). Alternative approaches with a similar
philosophy have been developed by other authors (e.g., van den Bosch 2000; Naab \& Ostriker 2006; 
Stringer \& Benson 2007; Dutton \& van den Bosch 2008). As a result of all these approaches, 
the models show that discs evolve inside out, and the predicted baryonic, stellar and luminous 
present-day scaling relations tend to be in agreement with observations (see e.g., FA00; 
Avila-Reese et al. 2008; Dutton \& van den Bosch 2008). Here we will explore 
the {\it direct predictions of such models regarding disk size evolution}. With this aim in mind, we will 
not include the possible effects of intermediate (uncertain) astrophysical processes 
[e.g., mass outflows, long time gas cooling, active galactic nuclei (AGN) feedback].

 In Section 2 a brief description of the galaxy evolutionary models 
and the strategy followed in this paper are presented. The results regarding disc size 
evolution and comparison with observational inferences are given in Section 3. Finally, a 
summary and the conclusions of the paper are presented in Section 4. We assume a concordance 
cosmology with $h=0.7$, $\Omega_{\Lambda}=0.7$, $\Omega_m=0.3$, $\Omega_b=0.04$, 
and $\sigma_8=0.8$.

\section{The model}

Our models calculate the formation and evolution of disc galaxies inside growing 
$\Lambda$CDM haloes, attempting to keep the underlying physics as transparent as 
possible.  The main physical ingredients are as follows (see for details FA2000; 
Avila-Reese \& Firmani 2000). An extended Press--Schechter approach is used to 
generate the halo mass accretion histories (MAHs) from the primordial Gaussian 
density fluctuation field. A generalized secondary infall model with elliptical 
orbits is applied to calculate the time--by--time 
virialization of the accreting mass shells (Avila-Reese et al. 1998).
The orbit ellipticity is fixed in such a way that the 
structure of the \LCDM\ haloes agrees well with results from cosmological N--body simulations 
(Avila-Reese et al. 1999; FA2000). 

A constant (baryon) fraction \fb~ of the mass of each shell is assumed to cool down 
in a dynamical time and form a disc layer. The accreting mass shells at the time of 
their virialization, $t_v$, are assumed to rotate rigidly  
with a specific angular momentum calculated as $j_{sh}(t_v)=\Delta J_h/\Delta M_h$, 
where $\Delta$ represents a difference between two time steps, and 
$J_h=\lambda _h GM_h^{5/2}/\left| E_h\right| ^{1/2}$, $M_h$, and $E_h$, are the halo 
total angular momentum, mass, and energy respectively; $\lambda _h$ is the halo spin 
parameter, assumed to be {\it constant in time}.
As a result of the assembling of these mass shells, a present-day 
halo ends with an angular momentum distribution close to the (universal) distribution measured 
in N-body simulations (Bullock et al. 2001). The radial mass distribution of the layer is 
calculated by equating its specific angular momentum to that 
of its final circular orbit in centrifugal equilibrium (detailed angular momentum conservation). 
The superposition of these layers form a gaseous disc, which tends to be steeper in the centre and 
flatter at the periphery than the exponential law. The gravitational interaction of disc and inner
halo during their assembly is calculated using the adiabatic invariance formalism (Blumenthal et al.
1986).


\begin{figure}
\vskip -0.8 true cm
\hskip -0.6 true cm
\psfig{file=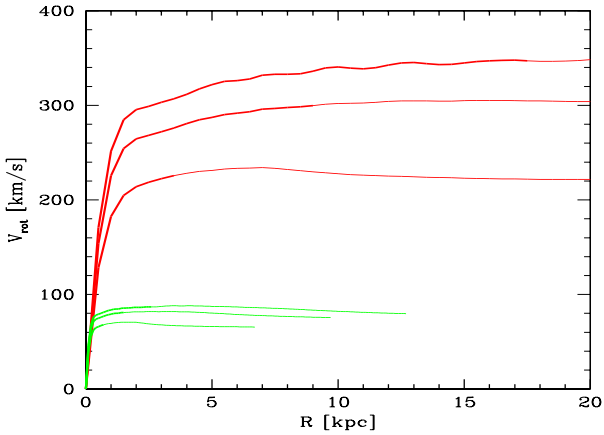 ,height=7.5 cm,width=10 cm} 
\caption
{Total rotation curves for "central" models of $\mh=7 \ 10^{10}\msun$ (bottom green) and 
$7 \ 10^{12}\msun$ (top red).
From top to bottom, the curves correspond to $z=$0, 1, and 3. 
Regions smaller than twice the effective radius (approximately one optical radius) are shown with 
the thick trace.
}
\label{evolrot}
\end{figure}


\begin{figure}
\hskip -3.6 true cm
\psfig{file=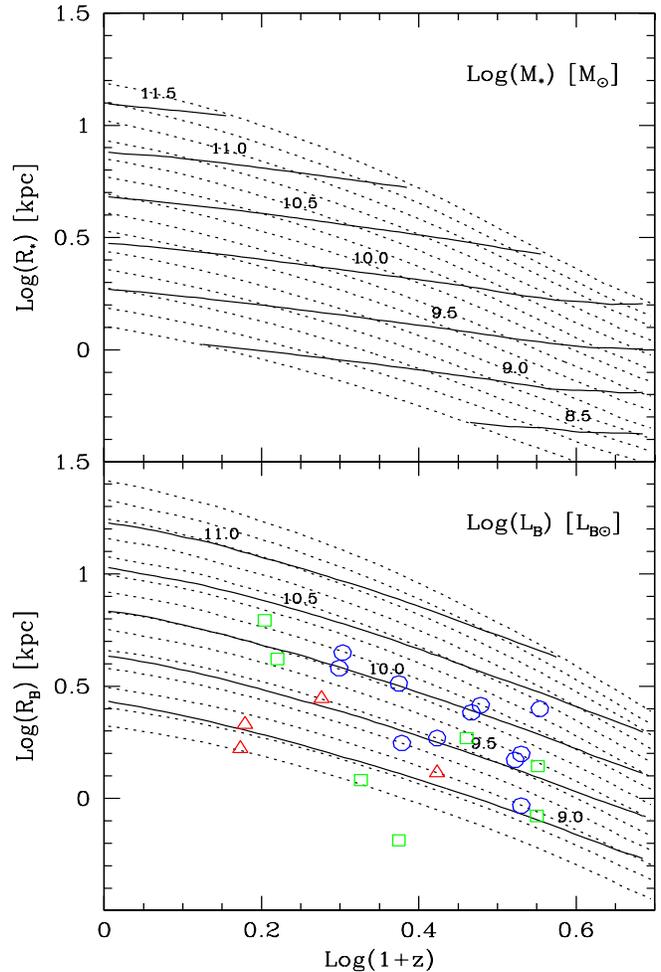,height=13.5cm,width=15cm} 
\caption
{Dotted curves are the \hs\ (upper panel) and \hB\ (lower panel) evolutionary paths of galaxy models 
that at $z=0$ have halo virial masses  (from top to bottom) \mh/\msun = $2.8 \ 10^{13}$,  $7.0\ 10^{12}$, 
$4.4\ 10^{12}$, $2.8\ 10^{12}$, $1.76\ 10^{12}$, $1.1\ 10^{12}$, $7.0\ 10^{11}$, 
$4.4\ 10^{11}$, $2.8\ 10^{11}$, $1.76\ 10^{11}$, $1.1\ 10^{11}$, and $7.0\ 10^{10}$ (logarithmic steps of 0.2). 
The solid curves connect points of constant Log\ms\ and Log$L_B$, respectively along the individual models. 
The corresponding constant values are indicated inside the panels; the unities
 are \msun\ for \ms\ (upper panel), and $L_{B\odot}$ for $L_B$ (lower panel).  Symbols in the 
 lower panel correspond to observed galaxies (Tamm \& Tenjes 2006): Log$L_B<$9.5 
(red triangles), $9.5\leq$Log$L_B\leq 10.0$ (green squares), and Log$L_B>$ 10.0 
(blue circles).
}
\label{radevol}
\end{figure}


A further step is the calculation of the {\it stellar} surface density profile. 
The previously mentioned processes and the fact that star formation (SF) is less efficient at 
the periphery than in the centre, produces stellar discs with a nearly exponential surface density 
distribution, and the size mainly determined by $\lambda_h$. 
A cusp is present in the inner region, where the bulge is identified in observed galaxies.
The disc SF at a given radius (assuming azimuthal symmetry) is triggered by the Toomre gas 
gravitational instability criterion and self--regulated by a balance between the energy input due 
to SNe and the turbulent energy dissipation in the ISM. 
Both such ingredients determine the gas disc height and the SF rate.
This physical prescription naturally yields a Schmidt-Kennicutt-like law. 
The SF efficiency depends on the gas surface density determined mainly by $\lambda_h$, and on 
the gas infall history, which is assumed to be proportional to the halo MAH.  
Very simple population synthesis models are used to calculate $B-$ and $V-$band luminosities
from the SF rates.

A given galaxy model is defined then by the halo MAH and $\lambda_h$, and by \fb.
The idea is {\it to keep the model as simple as possible in order to explore the direct consequences
of the \LCDM\ cosmogony on disc galaxy evolution}. 
Since we are interested here in the generic evolutionary trends, we study 
only the ``central'' models of different masses characterized by:
\begin{itemize}
\item  the averaged MAH corresponding to the given halo mass \mh, 
\item a value of $\lambda_h=0.035$, which is close to the mean of relaxed haloes measured in numerical 
simulations (e.g., Bett et al. 2007), 
\item an average value of $\fb=0.03$
\end{itemize}
Note that all these physical ingredients have in fact dispersions (not taken into account here) 
that, of course, will produce a wide scatter in the values of the output galaxy properties. 

A remark has to be made about $\fb$. Several pieces of evidence suggest values of 
$\fb \lsim 0.05$, though \fb\ is expected to vary with mass, being smaller for both very 
low and very high luminosity galaxies. Some physical mechanisms suggested for producing 
such low values are SN-driven galaxy mass outflows, long gas cooling times in massive haloes, 
SN and AGN halo gas re-heating. Given the current uncertainties in the understanding of these processes 
as well as in the observational determination of the halo masses, \mh, we  have preferred 
here to use a constant value for \fb\ equal to 0.03. With such a value, our disc galaxy models 
agree with observed dynamical properties, mainly the approximate flatness of the rotation 
curves at different epochs and the present-day disc-to-total velocity ratios at the 
maximum of the rotation curve (see e.g., Mo et al. 1998; FA2000; Zavala et al. 2003; 
Gnedin et al. 2007; Dutton et al. 2007).  

In Fig. \ref{evolrot} we plot the total rotation curve evolution ($z=0, 1$ and 3) corresponding 
to our ``central'' models with $\mh=7\ 10^{10}\msun$ (bottom green) and $7\ 10^{12}\msun$ (top red). 
The curves are nearly flat around twice the effective stellar radius \hs\ (indicated by
the thick trace), where \hs\ is the radius where half of the total stellar mass \ms\ is 
contained; two \hs\ correspond approximately to one optical radius.

\begin{figure}
\hskip -3.6 true cm
\psfig{file=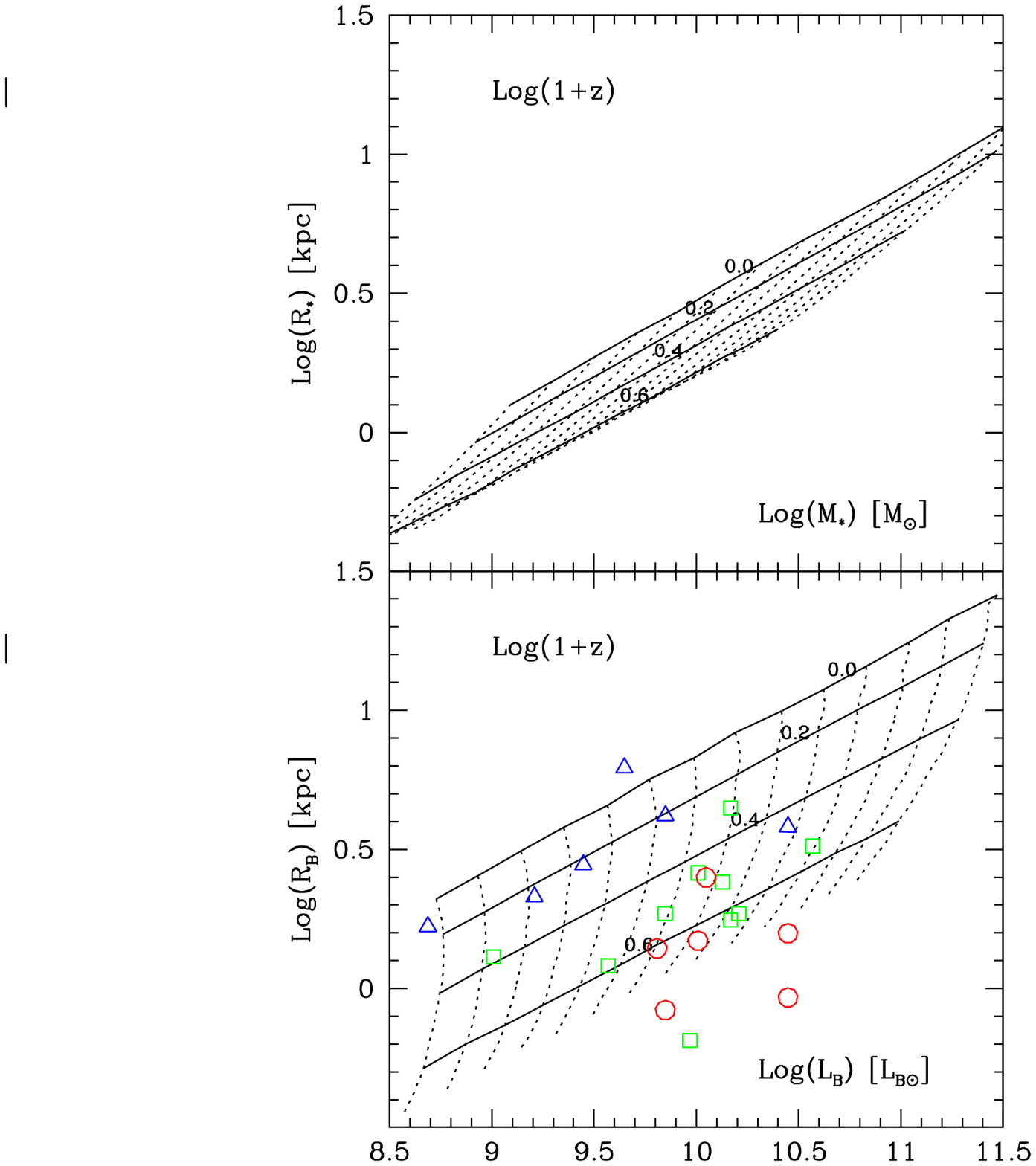,height=13.5cm,width=15cm} 
\caption
{Relations \hs--\ms\ (upper panel) and \hB--$L_B$ (lower panel) at four epochs, indicated 
inside the panels as Log(1+$z$), for the "central" model galaxies studied here. Dashed
lines are the evolutionary tracks of individual models. Symbols in the  lower panel 
correspond to observed galaxies (Tamm \& Tenjes 2006): Log(1+$z$)$< 0.3$ (blue triangles), 
$0.3\leq $Log(1+$z$)$\leq 0.5$ (green squares), and Log(1+$z$)$> 0.5$ (red circles)
}
\label{MLevol}
\end{figure}

\section{Results}

In Fig. \ref{radevol}, the dotted lines show the evolution of the stellar effective 
radius, \hs\ (upper panel), and the $B-$band effective radius, \hB\ (lower panel) 
for models with total present-day virial masses ranging from $\mh=7 \ 10^{10}\msun$ to 
$\mh=2.8 \ 10^{13}\msun$ by steps of 0.2 in logarithm. Both \hs\ and \hB\ significantly 
decrease with $z$, the decreasing being slightly steeper for the more massive galaxies. 
For present-day disc stellar masses ranging from $\approx 10^9$ 
to $\approx 10^{11} \msun$, the stellar effective radii decreases up to
$z=1$ by factors from 1.75 to 1.9, respectively; up to $z=2$, these
factors are from 2.8 to 3.2, respectively. In the $B$ band, for the same models, 
which at $z=0$ have $L_B$ from $\approx 4.2 \ 10^{10}$ to $\approx 5.3 \ 10^{8}\ L_{B\odot}$, 
the corresponding radius decreasing factors up to $z=1$ are from 1.75 to 1.85,
and up to $z=2$, are from 2.8 to 3.3, respectively.
The stellar and $B$-band radius evolutions tend to be similar on average, 
while \hB\ tends to be 1.4-1.6 times larger than \hs\ (at $z=0$).

Along each curve in Fig. \ref{radevol}, \ms\ and $L_B$ are also changing. 
We may connect the points in the different curves that have the same \ms\ 
value (upper panel) and $L_B$ value (lower panel). These curves (solid line) 
show the radius that disc galaxies of a fixed \ms\ ($L_B$) have at different 
epochs; {\it this is actually the kind of information that observations can provide, 
rather than the individual radius evolution}. 
The change of radius with $z$ along the \ms=constant curves is small. 
For $z<2.5$, \hs(\ms=constant) decreases roughly as $(1+z)^{-0.4}$ for all the masses 
studied here. In the case of the $L_B$=constant curves, the effective radius decrease with 
$z$ is strong and it resembles the decreasing law of the individual models. 
For $z\lesssim 0.75$, \hB($L_B$=constant) decreases roughly as $(1+z)^{-0.85}$, 
and for $0.75\lesssim\ z\ \lesssim 3$ it does as $(1+z)^{-1.1}$. 

The different radius dependences on $z$ for models in the upper and lower 
panels of Fig. \ref{radevol} are explained basically by the differences in 
the evolution of \ms\ and $L_B$, which implies a significant change with $z$ 
of the \ms/$L_B$ ratio. These different evolutionary paths can also be 
appreciated in the $\hs(z)$ vs $\ms(z)$ and $\hB(z)$ vs $L_B(z)$ diagrams 
of Fig. \ref{MLevol} (upper and lower panels), respectively, where the solid 
lines connect models at a given $z$, while the dashed lines show the 
evolutionary tracks of individual models. At least for the ``central'' models 
studied here, we see that at $z=0$, $\hs\propto \ms^{0.38}$ 
and $\hB\propto L_B^{0.40}$. Furthermore, the slopes of these correlations 
roughly remain the same at higher redshifts in such a way that the changes 
with $z$ of the relation zero-points, $\hs(z)/\ms(z)^{0.38}$ and $\hB(z)/L_B(z)^{0.40}$, 
describe basically the curves of radius change at \ms=constant and $L_B$=constant. 
plotted in Fig. \ref{radevol}. Thus, our models show that using the radius change 
with $z$ for a fixed mass (luminosity), or using the zero-point evolution of 
the radius-mass (-luminosity) relations for inferring the size evolution of 
galaxies, depends on how $\ms$ and $L_B$ do evolve. 

Our results agree conceptually with those by Somerville et al. (2008). 
However, the predictions of these authors refer to 'static' populations and to baryonic 
discs rather than to individual evolving stellar and luminous discs (their approach 
does not allow to calculate gas transformation into stars). 
With our evolutionary models, we have generated a plot similar to the upper panel 
of Fig. \ref{radevol} but for the baryonic effective radius and for curves of \md=constant. 
The obtained curves follow very closely the stellar ones, showing that the radius evolution 
of the baryonic and stellar disc is in fact quite similar.

\subsection{Model results vs observational inferences}

In the lower panels of Figs. \ref{radevol} and \ref{MLevol}, the data corresponding to 22 disc 
galaxies with photometric $z'$s between 0.5 and 2.6 reported in Tamm \& Tenjes (2006) are 
plotted. The authors used deep Near Infrared Camera and Multi-Object Spectrometer (NICMOS) 
$J$ and $H$ band images from the HDF-S field to infer rest--frame $B$-band photometric 
properties. The sample is not complete in any sense but it 
gives an idea of the typical $B$-band radii and luminosities of normal galaxies at different 
redshifts. In order to compare with the models, in Fig. \ref{radevol} the sample is split in 
three luminosity ranges indicated in the figure caption.
The luminosities were calculated from the absolute magnitudes reported in the paper, 
and both $L_B$ and \hB\ were rescaled to $h=0.7$ (they used $h=0.65$).

In Fig. \ref{MLevol}, the sample is split in three $z$ ranges indicated in the figure caption.
As seen in Figs. \ref{radevol} and \ref{MLevol}, the few observed galaxies 
seem to be consistent, within the large scatter, with the radius evolution of our models. 
If any, the observed galaxies at a given $z$ and radius, seem to be on average brighter in 
the $B$ band than models, specially for higher redshifts. Since $L_B$ traces recent SF rate, 
such a result may be related to the so--called downsizing problem. Such difficulty 
appears also in the similarity between the slopes of the \ms-\hs\ and $L_B$-\hB\ relations 
showed by the models. From observations, this slope is significantly steeper for 
the $L_B$-\hB\ relation than for the \ms-\hs\ one (e.g., Avila-Reese et al. 2008), 
reflecting this basically the fact that our models do not reproduce the observed correlation 
between color and mass or luminosity. While the predicted slope in the $B$ band seems
to be only slightly shallower than observations, it is the slope of the \hs-\ms\ relation
that is steeper than the observational inferences presented in Avila-Reese et al. (2008;
see also Pizagno et al. 2005).
This difference is even more dramatic at lower masses when comparing models to the
observational results by Shen et al. (2003). In any case, we do not attempt to carry out
a more detailed comparison of models and observations in the \ms-\hs\ or $L_B$-\hB\ 
diagrams because these relations are actually highly scattered and segregated by SB
and morphological type (see e.g., Graham \& Worley 2008).

\begin{figure}
\hskip -3.6 true cm
\psfig{file=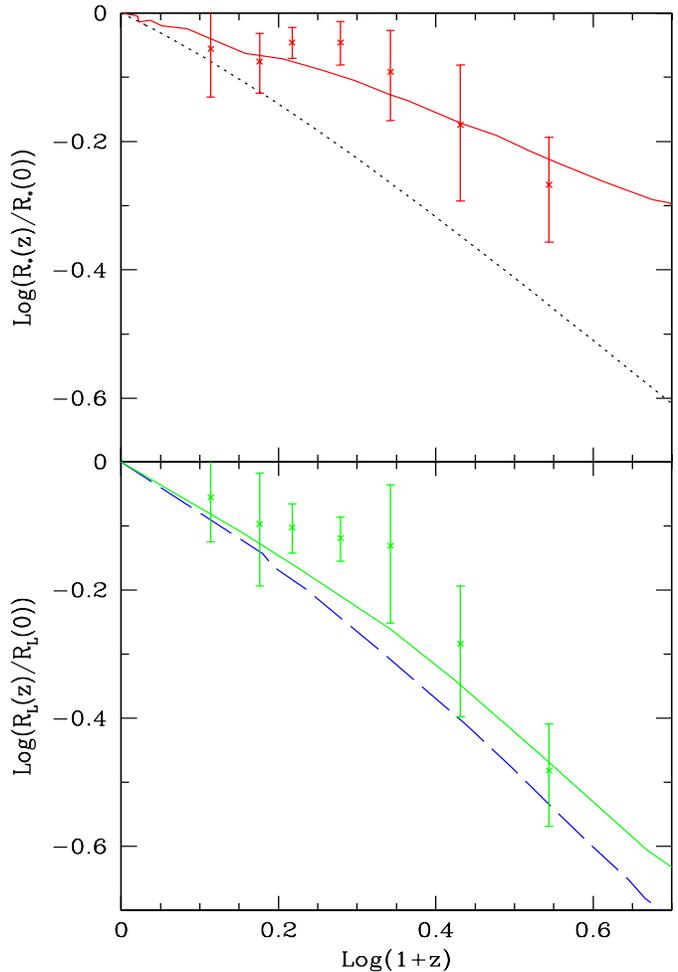,height=13.5cm,width=15cm} 
\caption
{{\it Upper panel:} Crosses are the mean stellar radii normalized to the present epoch for 
observed disc galaxies more massive at each $z$ than $\ms\approx 3\ 10^{10}h_{70}^{-2} \msun$
(from Trujillo et al. 2006).  The error bars are the 2$\sigma$ scatter on the mean 
values estimated from the Log[\hs($z$)/\hs(SDSS)] distribution.  The solid line is 
the \hs\ change normalized to $z=0$ of model galaxies that at each $z$ have 
$\ms = 3\ 10^{10}\msun$ (the behavior of the normalized \hs\ change for larger 
values of \ms=constant. is quite similar to the one plotted here). Because the models 
correspond only to the "central" cases, there is not any scatter to report. The dotted 
line is the normalized halo radius change with $z$ at a fixed \mh\ for all redshifts 
(eq. \ref{radvir}).  {\it Lower panel:} Same as in the upper panel but in the $V$ band, for galaxies 
more luminous at each $z$ than $L_V \approx 3.4 \ 10^{10}h_{70}^{-2} L_{V\odot}$  
(rest-frame). The  model galaxies are for $L_V = 3\ 10^{10} L_{V,\odot}$ (solid line) and
$L_B=  2\ 10^{10}  L_{B,\odot}$ (dashed line) at each $z$.
}
\label{radobs}
\end{figure}

In order to infer reliable estimates of the average disc radius change with $z$, samples 
at different redshift bins, complete (corrected by the selection effects) down to a given 
luminosity, are necessary.  The changes of the luminosity and mass-size relations in 
the range $1\lesssim z \lesssim 3$ of luminous (massive) field galaxy populations have 
been inferred recently from the Faint InfraRed Extragalactic Survey (FIRES) by Trujillo 
et al. (2006; see also Trujillo 
et al. 2004). Due to the large number of objects in their sample, the authors are able to apply 
a completeness function to three redshift bins.  Late-type galaxies are selected as those 
with the S\'ersic index $n\lesssim 2$. Combining the analysis of FIRES data with the 
results obtained by using the Galaxy Evolution from Morphologies and SEDs (GEMS) 
survey at $0.2<z<1$ (Barden et al. 2005) and tying both
to the $z\sim 0$ results from SDSS (Shen et al. 2003), Trujillo et al. (2006) presented a 
comprehensive picture of the change with $z$ of the mean radius and its dispersion 
for galaxies more luminous than 
$L_V \approx 3.4 \ 10^{10} L_{V\odot}$ (rest-frame) or more massive than 
$\ms\approx 3\ 10^{10} \msun$ at any epoch. Their results for the $n<2$ (late-type) 
galaxies are reproduced in the upper (mass) and lower (luminosity) panels of Fig. \ref{radobs}; 
the error bars are the 2$\sigma$ scatter on the mean values estimated from the 
Log$[R_{e}(z)/R_{e}(SDSS)]$ distribution.

The solid lines in Fig. \ref{radobs} show how the normalized stellar and V band effective 
disc radii change with $z$ for models that at any $z$ have $\ms=3\ 10^{10}\msun$ 
(upper panel) and rest-frame $L_V=3\ 10^{10} L_{V,\odot}$ (lower panel), respectively.
In fact, the curves are quite similar for \ms=constant in the range  
$10^{10}\msun-3\ 10^{11}\msun$ and $L_V$=constant. in the range 
$2\ 10^{10}L_{V,\odot}-10^{11}L_{V,\odot}$ (see Fig. \ref{radevol}). 
Thus, Fig. \ref{radobs} captures the typical shape of the normalized radius change 
with $z$ for galaxies of the same mass or luminosity at different $z'$s.
Some observations suggest that the normalized \hs\ change with $z$ could be slightly 
steeper for the largest masses (Barden et al. 2005; Trujillo et al. 2006; Franx et al. 2008). 
However, any possible dependence with mass is within the large scatter of the \hs-\ms\ 
correlations at different $z'$s and the observational uncertainties. 
Besides, at higher $z'$s the samples are probably biased to observe earlier-type 
galaxies since they strongly emit in UV at their early formation phases; and 
early-type galaxies are more compact than late-type galaxies. 

The dotted line in the upper panel shows the redshfit change of the normalized halo 
virial radius, \rh\ given by eq. (\ref{radvir}). Our evolutionary models show that the 
relative radius change with $z$ for a fixed disc stellar mass is much slower than in 
the case of dark matter haloes. For completeness, we also plot in the lower panel 
of Fig. \ref{radobs}, models in the $B$ band, for rest-frame $L_B=  2\ 10^{10} L_{B,\odot}$ 
(dashed line). The sequence is clear: the decrease of the normalized radius with $z$ 
for a fixed luminosity (or stellar mass, which is closely related to a $K$ band) 
is steeper as the pass-band is bluer.

We did not include in our models the possible cooling flow regime for gas infall due to 
cooling times longer than the dynamical time in massive haloes. By truncating the late 
infall of high-angular momentum gas, the disc radius increasing is expected to be halted. 
Detailed calculations (e.g., Dutton \& van den Bosch 2008) show that gas infall is 
truncated at $z\approx 0$ for haloes of virial mass $\mh \approx 6\ 10^{11}\msun$ 
(corresponding to $\ms\approx 10^{10}\msun$ and $L_V\approx 7\ 10^{9} L_{V\odot}$ in 
the case of the models presented here). As the halo is more massive than this value, 
the truncation of gas infall starts earlier. 
Therefore, the long cooling time in massive haloes is expected to work in the direction of 
flattening the late radius change with $z$ of galaxies of masses $\ms > 10^{10}\msun$ 
(or $L_V> 7\ 10^{9} L_{V\odot}$), approaching even more models to observations in 
Fig. \ref{radobs}.
We did not include either the possible feedback mass outflow mechanism, which introduces a
decrease of \fb\ as \mh\ is smaller. Such a dependence is expected to produce a 
flattening of the \hs-\ms\ relation at lower masses (Shen et al. 2003; Somerville et al. 2008),
improving perhaps the comparison with observations in that regards this relation.

\section{Conclusions}

We have presented results on disc size change with redshift obtained by means  
of galaxy {\it evolutionary} models within the context of the \LCDM\ cosmogony. 
Under the assumption of detailed angular momentum conservation, discs are formed in 
centrifugal equilibrium inside growing \LCDM\ virialized haloes, taking into account 
the central halo (adiabatic) contraction due to the disc gravitational drag. 
The gas infall rate is assumed to be a constant fraction \fb\ of the halo mass 
accretion rate. The hydrodynamic equations for the gaseous and stellar discs are 
resolved coupled under the assumption of azimuthal symmetry and vertical virial 
equilibrium. The SF is triggered by the Toomre instability criterion and self-regulated 
by an energy balance between SN energy input and ISM turbulent dissipation. 
The driving parameters of the models are the halo MAH, the spin parameter $\lambda_h$, 
and the mass galaxy fraction, \fb. 

Since our main goal was to explore the general behavior of disc size evolution, here
we took into account only ``central'' models constructed with the average MAH 
corresponding to a given \mh, a value of $\lambda_h=0.035$ close to its mean, and a 
global average value for \fb=0.03; both $\lambda_h$ and \fb\ are further assumed to 
be constant in time. {\it We avoided the introduction of some (uncertain) astrophysical processes 
in order to explore in a transparent fashion the direct consequences of the \LCDM\ 
cosmogony on disc size (and mass/luminosity) evolution}; the assumed value of \fb\ 
is based on the success of the models to produce nearly flat rotation curves and
realistic disc-to-halo velocity ratios.  Our main conclusions are:

- The model stellar and $B$ luminosity  disc effective radii significantly decrease with $z$ 
(inside-out disc formation), the decreasing at high $z'$s being slightly steeper for the more 
massive galaxies (Fig. \ref{radevol}).  The radius growth in the $B$ band is only slightly 
steeper than the one of the stellar disc, and at $z=0$, \hB\ is 1.4-1.6 times larger than \hs. 

- The change of \hs\ with $z$ at a fixed \ms\ --this is the kind of information able to infer 
from the observations-- is much weaker than the radius evolution of the individual models 
(Fig. \ref{radevol}). 
For $z\leq 2.5$, \hs(\ms=const)$\propto (1+z)^{-0.4}$ for all the masses studied here.
This law is much shallower than the one corresponding to the haloes, which was widely 
used in the literature for comparisons with observations.
On the other hand, in the $B$ band, the decrease of \hB($L_B$=const) with $z$ is strong
and it resembles the radius decreasing law of the individual models. For $z\lesssim 0.75$, 
\hB($L_B$=const)$\propto (1+z)^{-0.85}$, and for $0.75\lesssim z \lesssim 3$, 
\hB($L_B$=const)$\propto (1+z)^{-1.1}$. Since the radius growth of the stellar 
and $B$ band discs is similar for the individual models, the difference in the radius 
change with $z$ for models of the same \ms\ and $L_B$ at all $z'$s is basically due 
to the differences in the individual evolutionary paths of \ms\ and $L_B$

- The slopes of the \hs-\ms\ and \hB-$L_B$ relations remain practically the same since $z=0$ 
up to $z=3$; these slopes at $z=0$ are $0.38$ and 0.42, respectively (Fig. \ref{MLevol}). 
Therefore, the changes with $z$ of the relations zero--points, $\hs(z)/\ms(z)^{0.38}$ and 
$\hB(z)/L_B(z)^{0.40}$, describe basically the radius change curves at \ms=constant and 
$L_B=$constant
shown in Fig. \ref{radevol}. The similarity in the slopes of both relations disagrees with 
observational inferences at $z=0$, which show a significant steepening from the stellar to 
the $B$ band relation. Such a behavior is related to the well known color-magnitude (mass) 
relation, which our models do not reproduce.

- Model predictions are in reasonable agreement with observational 
inferences of the typical \hs\ and \hV\ change with $z$ for disc galaxies more massive than 
$3\ 10^{10} \msun$ and more luminous than $3.4 \ 10^{10} L_{V\odot}$ 
(rest-frame) at any $z$, respectively (Fig. \ref{radobs}). These large limits in \ms\ 
and $L_V$ are dictated by the completeness functions required to introduce at different 
redshifts in order to eliminate the selection effects (Trujillo et al. 2006). In more detail, 
the predicted radius decrease with $z$ for massive galaxies tends to be steeper than the one 
inferred from observations up to $z\approx 1$, specially in the $V$ band; for higher redshifts 
(up to the observational limit of $z=2.5-3$), models and observations agree well. 

- Unfortunately, there are not yet available complete high-redshift galaxy samples that would
allow to infer the typical stellar and luminous disc size evolution of sub-$L_*$ galaxies. 
Within the large scatters, our models seem to be consistent with small (not complete) 
samples of sub-$L_*$ disc galaxies with available rest-frame photometric properties 
(e.g., Tamm \& Tenjes 2006) concerning the measured values of \hB\ and $L_B$ at different 
$z'$s. 
 
The kind of model predictions presented here, unlike previous ones made only on the 
level of galaxy populations at different redshifts, allow to understand the evolution of 
individual galaxies and its connection to the properties of the overall galaxy populations. 
The model predictions are roughly consistent with the currently available observational 
information on disc size evolution, showing that such an evolution is driven mainly by the 
\LCDM\ halo evolution. The role of some astrophysical mechanisms in play (e.g, mass outflows, 
late gas infall truncation in massive haloes, stellar and AGN gas re-heating) seems to be
of less relevance for the disc size evolution. More observational inferences
are required in order to constrain different models.

\section*{Acknowledgments}

We are grateful to the referee for useful comments that have helped to improve the
manuscript. V.A.R. thanks PAPIIT-UNAM grant IN114509 and CONACyT grant 60354 for partial funding.

\end{document}